\documentclass[sigconf, authorversion, nonacm]{acmart} 

\usepackage{multirow}

\usepackage{soul}

\usepackage{listings}
\usepackage{xcolor}

\usepackage{enumitem} 

\definecolor{codegray}{rgb}{0.5,0.5,0.5}
\definecolor{codegreen}{rgb}{0,0.6,0}
\definecolor{codeblue}{rgb}{0,0,1}

\lstdefinestyle{customcpp}{
  language=C++,
  basicstyle=\ttfamily\footnotesize,
  keywordstyle=\color{codeblue}\bfseries,
  commentstyle=\color{codegreen},
  numberstyle=\tiny\color{codegray},
  numbers=left,
  stepnumber=1,
  breaklines=true,
  captionpos=b,
  frame=single
}

\AtBeginDocument{%
  }

\begin{document}

\title{P4sim: Programming Protocol-independent \\ Packet Processors in ns-3}

\author{Mingyu Ma}
\email{mingyu.ma@tu-dresden.de}
\orcid{0009-0005-7045-4096}
\authornotemark[1]
\affiliation{%
  \institution{Centre for Tactile Internet with Human-in-the-Loop (CeTI), Dresden University of Technology}
  \city{Dresden}
  \state{Saxony}
  \country{Germany}
}


\author{Giang T. Nguyen}
\email{giang.nguyen@tu-dresden.de}
\orcid{}
\authornotemark[1]
\affiliation{%
  \institution{Centre for Tactile Internet with Human-in-the-Loop (CeTI), Dresden University of Technology, Germany}
  \city{Dresden}
  \state{Saxony}
  \country{Germany}
}

\renewcommand{\shortauthors}{Ma et al.}

\begin{abstract}

Programmable data planes enable users to design data plane algorithms for network devices, providing extensive flexibility for network customization. 
Programming Protocol-Independent Packet Processors (P4) has become the most widely adopted abstraction, programming language, and framework for data plane programming. However, existing simulation platforms lack high-performance support for P4-based networks.
This paper introduces P4sim, a high-performance P4-driven simulation framework built on bmv2 and NS4, seamlessly integrated with ns-3. It improves queue modeling, time scheduling, and P4 architecture support, extending compatibility to V1model, PSA, and PNA. P4sim enables efficient packet processing, accurate time tracking, and seamless interaction between P4-enabled hosts and switches.
We evaluate the P4sim in terms of performance and queue management and demonstrate its capabilities using two common use cases: Basic Tunneling and Load Balancing. The results highlight the P4sim as a powerful tool for advancing research and education in programmable networks.

\end{abstract}




\maketitle

\section{Introduction}
\label{sec:introduction}

Modern computer networks, including local area networks, large-scale structured networks, and data center backbones, have grown increasingly complex, requiring in-depth research for effective analysis and optimization. Among various methodologies, simulation has emerged as a critical tool for evaluating design quality and network performance by capturing key aspects such as scalability, security overhead, and resource efficiency \cite{campanile2020computer}.
In particular, simulating network devices demands precise behavioral modeling to ensure realistic and reliable evaluations. By incorporating detailed behavioral models, researchers can analyze device interactions, protocol performance, and overall network efficiency in a controlled environment. 
Researchers integrate these models into the target network within the simulator to extensively test and validate their correctness, feasibility, and performance under diverse conditions.

With the rise of programmable data planes over the past decade, Programming Protocol-independent Packet Processors (P4) has emerged as a powerful language for defining customized packet-processing pipelines in modern networks \cite{bosshart2014p4}. Simulating P4-based network devices requires an accurate P4 behavioral model to represent their parsing, processing, and forwarding logic. Behavioral model version 2 (bmv2) \cite{p4lang/behavioral-model} is the second-generation reference software switch developed by the P4 Language Organization. With its well-developed functionality and robust implementation, bmv2 provides a comprehensive environment for the development and experimentation of P4-based network functions. 
Realistic network simulations require more than a switch model; traffic sources, links, and channels must also be included.

Network simulator version 3 (ns-3) is a widely used open-source network simulation environment based on discrete-event simulation. Designed to be modular, extensible, and programmable, ns-3 provides a flexible platform for evaluating various network protocols and architectures. By integrating P4 behavioral models into ns-3, the NS4 \cite{fan2017ns4p4simulator, bai2018ns4p4simulator} enables performance, scalability, and correctness evaluation of programmable network designs in a controlled simulation environment. However, the existing NS4 has limitations in accurately modeling advanced P4 behaviors and integrating efficiently with ns-3's simulation architecture.
To overcome these challenges, we introduce the P4sim\footnote{\url{https://github.com/HapCommSys/p4sim}} module in ns-3, which enhances switch modeling, time representation, and extensibility. Specifically, this work makes the following key contributions:

\textbf{R1: High performance with time representation:} Implemented queue modeling within switches to enable high-speed link simulation, ensuring that each switch processes one packet per event for improved efficiency.

\textbf{R2: Extended P4 Support:} Enhanced compatibility with modern P4 architectures: V1model, Portable Switch Architecture (PSA) \cite{p4lang/psa}, Portable NIC Architecture (PNA) \cite{p4lang/pna}, enabling support for additional P4 constructs and more complex pipeline behaviors.

\textbf{R3: Protocol Independence:} Expanded flexibility through P4 custom header design, ensuring both P4 switches and hosts remain protocol-agnostic for broader simulation use cases. 

\textbf{R4: Seamless ns-3 Integration:} Strengthened interoperability with existing ns-3 modules, follows modular design and supports \textit{Waf} and \textit{Cmake} builds.

The rest of the paper is organized as follows. 
Section~\ref{sec:related_work} summarizes and discusses related work on P4 platforms and ns-3-based P4 implementations, and outlines the key challenges.
Section~\ref{sec:approach} provides a detailed overview of bmv2 and NS4, which form the foundation of our proposed P4sim.
Section~\ref{sec:ns3p4simulator} presents the design of P4sim, including its switch model, time scheduling mechanism, and custom header format. 
Next, we present the module implementation in Section \ref{sec:moduleimplementation} and performance and model evaluation in Section \ref{sec:evaluation}. 
Finally, Section \ref{sec:conclusion} wraps up the paper with conclusions and future directions. 



\section{Related Work}
\label{sec:related_work}

Several studies have explored P4-based programmable networks across different platforms, including hardware-based solutions such as Tofino switches \cite{tofinoPlatform}, FPGA-based accelerators, and SmartNICs, as well as software-based platforms like bmv2. 
Each of these platforms is designed for different use cases. 
Tofino  is primarily used for production-level deployments. 
Net-FPGA \cite{zilberman2014netfpga} is tailored for hardware acceleration, enabling researchers to develop and test high-speed packet processing pipelines. 
T4P4S \cite{voros2018t4p4s} focuses on high-performance packet processing, making it suitable for scenarios requiring optimized, low-latency data plane implementations. 
bmv2 is mainly used for learning and debugging, allowing researchers and developers to experiment with P4 programs in an emulation environment. bmv2 has performance limitations \cite{p4lang/bmv2/performance} as it processes packets in real machine time, a drawback that becomes more evident when used with Mininet \cite{p4lang/tutorials}.
Given the hardware dependencies and performance constraints of existing platforms, a high-performance, hardware-independent simulator is essential to facilitate scalable and accurate P4 network research.


Building on this, we conducted a detailed analysis of various software-based P4 implementations, summarized in Table \ref{tab:p4_software_comparison}. Our analysis focuses on three ns-3-based simulation modules, ns3-bmv2 \cite{PIFO-TM/ns3-bmv2}, NetDevices \cite{mrosan/P4SwitchNetDevice}, NS4 \cite{fan2017ns4p4simulator, bai2018ns4p4simulator} and P4sim, which represent different approaches to integrating P4 programmability within ns-3. 
ns3-bmv2 implements the P4 module within the traffic control module of ns-3, allowing it to be deployed along a single network path. Although this approach emphasizes packet queueing and scheduling mechanisms, it largely overlooks core P4 functionalities such as routing and switch-level processing. This makes ns3-bmv2 more suitable for traffic shaping than full-fledged P4-based network simulations. 
NetDevices takes a different approach by utilizing T4P4S, a re-targettable P4 compiler, to convert P4 programs into C code, which is then incorporated into ns-3 simulation scripts. While this method enables P4-based switch modeling within ns-3, it comes with significant complexity in terms of integration, debugging, and traceability. The generated code requires manual modifications and lacks a unified management framework within the ns-3 modular structure, making it challenging to maintain and extend. 
NS4 is a P4-driven network simulator, it simplifies the development of P4 behavioral models and represents the first research effort to simulate P4-enabled networks. It integrates P4 packet processing and forwarding by linking the bmv2 library. However, it lacks support for queues, simulation virtual time, and primitive actions, and the functionality of the switch model is largely overlooked. 
A comparison of existing methods and software-based platforms reveals that none fully satisfy the R1-R4 requirements outlined in Section \ref{sec:introduction}, particularly in terms of high performance with switch time representation. 
To realize a solution that fulfills all aspects of R1–R4, it is essential to overcome several technical challenges. The following outlines the main challenges encountered during the design and implementation of our approach.

\begin{table*}[]
\centering
\begin{tabular}{l|ccccccc}
\hline
\multirow{2}{*}{\shortstack{\textbf{Software-Based} \\ \textbf{Platforms}}} & \textbf{bmv2 } & \textbf{p4c-dpdk} \cite{p4lang/p4dpdktarget} & \textbf{p4c-ebpf} \cite{p4lang/p4c/backends/ebpf} & \textbf{ns3-bmv2} & \textbf{NetDevices} & \textbf{NS4} & \textbf{P4sim} \\ 
 & \cite{p4lang/behavioral-model} & \textbf{/ t4p4s} \cite{voros2018t4p4s} & & \cite{PIFO-TM/ns3-bmv2} & \cite{mrosan/P4SwitchNetDevice} & \cite{fan2017ns4p4simulator, bai2018ns4p4simulator} & \\ \hline
Time Type & real time & real time & real time & virtual time & real time & virtual time & virtual time \\ 
Time Model \cite{martinek2022white} & $\checkmark$ & $-$ & $-$ & \texttimes & \texttimes & \texttimes & $\checkmark$ \\ 
P4 Functional Support & $\checkmark$ & $\checkmark$ & $\checkmark$ & $\diamond$ & $\diamond$ & $\diamond$ & $\checkmark$ \\ 
P4 Version 16 and 14 & $\checkmark$ & $\diamond$ & $\diamond$ & $\checkmark$ & $\checkmark$ & $\checkmark$ & $\checkmark$ \\ 
Performance & $\diamond$ & $\checkmark$ & $\checkmark$ & $\checkmark$ & $\diamond$ & $\checkmark$ & $\checkmark$ \\ 
Tracing \& Logging & $\checkmark$ & $\diamond$ & $\checkmark$ & $\checkmark$ & \texttimes & \texttimes & $\checkmark$ \\ 
\shortstack{P4 Switch \\ Architecture} & 
\shortstack{V1model, \\ PSA, PNA} & 
\shortstack{PSA, \\ PNA} & 
\shortstack{PSA, \\ eBPF backend} & 
\shortstack{Incomplete \\ V1model} & 
\shortstack{V1model, \\ PSA} & 
\shortstack{Incomplete \\ V1model} & 
\shortstack{V1model, \\ PSA, PNA} \\ \hline
\end{tabular}
\caption{Comparison of software-based P4 platforms.}
\label{tab:p4_software_comparison}
\end{table*}

\textbf{C1: Lack of Queue Management and Scheduling in Switches.} Current implementations of NS4 and related works lack a built-in queueing model for switches, which restricts their ability to support essential mechanisms such as scheduling, buffering, and queue management. 

\textbf{C2: Missing Virtual Time Representation.} Virtual time is fundamental for accurate network simulation, as it enables precise modeling of packet transmission, delays, and scheduling across different network components \cite{jefferson1985virtual}. ns-3 relies on virtual time to synchronize events across nodes, but existing switch models lack proper time representation. Without that, switches cannot accurately perform essential functions such as monitoring and congestion control. 

\textbf{C3: Limited Support for Modern P4 Architectures.}
NS4 and other solutions primarily integrate bmv2's V1model but lack support for PSA and PNA architectures. This limitation reduces its applicability to real-world programmable networking research, as P4 typically targets PSA for deployment in actual P4 switch processing pipelines.

\textbf{C4: Lack of a Protocol-Independent Simulation Framework.} P4 provides a protocol-independent approach to defining and processing packet headers. However, existing ns-3 simulations lack a protocol-independent generalized framework.

\section{Approach}
\label{sec:approach}

Given the challenges C1–C4 discussed in Section~\ref{sec:related_work}, existing approaches still fall short in addressing them effectively. As a switch behavioral model, bmv2 allows for detailed simulation of packet processing within a switch, while NS4 extends ns-3 to support P4-based network modeling. Despite their limitations, these tools offer valuable insights and serve as important references for our work.
Our approach builds upon bmv2 and NS4, and the following sections provide a detailed introduction to P4 and bmv2, followed by an explanation of ns-3 and the NS4 extension.

\subsection{P4 and bmv2}
\label{subsec:p4andbmv2}
P4 is a domain-specific programming language designed to define and control protocol-independent packet forwarding behavior on programmable network devices. Unlike traditional fixed-function switches, which support only predefined protocols, P4 allows users to define custom header structures, specifying exact fields and their bit widths for extraction and modification. Packet forwarding or dropping can be determined based on calculations or memory lookups within the processing pipeline. This flexibility enables the implementation of new or proprietary protocols without requiring hardware changes, making P4 highly adaptable to evolving network requirements.

bmv2 (behavioral model version 2) is the second-generation reference software switch developed by the P4 language consortium \cite{p4lang/behavioral-model}.
From survey \cite{liatifis2023advancing} \cite{hauser2023survey}, bmv2 is the most widely used software-based switch implementation for P4. 
Network devices such as switches and NICs feature diverse hardware architectures, requiring different processing models to accurately reflect their behavior. bmv2 supports multiple models, including V1model, PSA, and PNA, to simulate their unique behaviors. V1model implements a basic match-action pipeline for programmable switches, it's originally introduced as the target for P4-16. PSA describes common capabilities of network switch devices and extends the v1model by incorporating features such as an ingress deparser and an egress parser. PNA is designed for high-performance packet processing on network interface cards (NICs), supporting applications in data center networking and in-network computing.
The \textit{simple\_switch} \cite{p4lang/simpleswitch} target is based on the V1model architecture, providing a conventional match-action pipeline, it's originally introduced as the target for P4-16, and primarily used in bmv2. The \textit{psa\_switch} \cite{p4lang/psa} target implements the PSA, which standardizes ingress, egress, and queuing mechanisms for both hardware and software switches. The \textit{pna\_nic} \cite{p4lang/pna} target is designed for high-performance NICs, enabling programmable packet processing directly on network adapters. 

\subsection{ns-3 and NS4}

ns-3 is a discrete-event network simulator for Internet systems, targeted primarily for research and educational use \cite{henderson2008network}. 
Currently, in ns-3 there is no dedicated Ethernet switch module. The closest existing component is \textit{ns3::BridgeNetDevice}, which serves as a virtual network device that connects multiple LAN segments. It operates more like a network hub rather than a traditional Ethernet switch, which transmits packets to the \textit{NetDevices} by checking the packet's header. Moreover, the bridge network device operates in a half-duplex mode, requiring mechanisms such as carrier sensing and collision detection, the same as those used in a Carrier Sense Multiple Access (CSMA) channel. This contrasts with full-duplex switching, where data transmission and reception can occur simultaneously without contention, as seen in Point-to-Point (P2P) channels. It can be viewed as equivalent to a full duplex link.

NS4 is a P4-driven network simulator supporting simulation of P4-enabled networks \cite{bai2018ns4p4simulator}. It has already integrated the V1model packet processing functions into ns-3, effectively deploying it with advanced capabilities such as flow tables configuration and programmable match actions. These components enable dynamic packet forwarding and in-network computing based on user-defined P4 scripts. The implementation of use cases like SilkRoad demonstrates its performance and potential.
The primary packet processing model in NS4 is the \textit{P4Model} and \textit{P4NetDevice}, which utilizes multiple CSMA \textit{NetDevices} as its ports. This design creates a shared communication environment that resembles a hub rather than an Ethernet switch.

\section{P4sim Design}
\label{sec:ns3p4simulator}

As introduced in Section \ref{sec:related_work}, we address the four identified challenges in this section. First, we tackle C1 through P4 switch modeling. Next, C2 is addressed by introducing a scheduling model with time representation. We then solve C3 by implementing different P4 architectures through a combined approach of switch and scheduling model. Finally, C4 is resolved by adding a custom P4 header.

\subsection{P4 Switch Model}
\label{sec:p4switchmodel}

In network simulations, switch modeling includes key aspects such as maximum packet forwarding rate, queue sizes, policies, intra-device latencies, and error models \cite{deviceindependentroutermodel}.
Our switch modeling will implement all these aspects, excluding the error model. 
The queue and processing policies are based on bmv2, with the packet processing unit simply inheriting from the external function \textit{bm::switch}.  Packet forwarding rates and intra-device latencies will be integrated into this unit using ns-3 scheduling and configurations.
Excluding the error model aligns with prior studies that emphasize switch behavior over link-layer error characteristics. Additionally, in high-speed wired networks, the impact of transmission errors is often negligible compared to queuing and scheduling delays, making error modeling less critical in this context.


\begin{figure*}[ht]
  \centering
  \includegraphics[width=\textwidth]{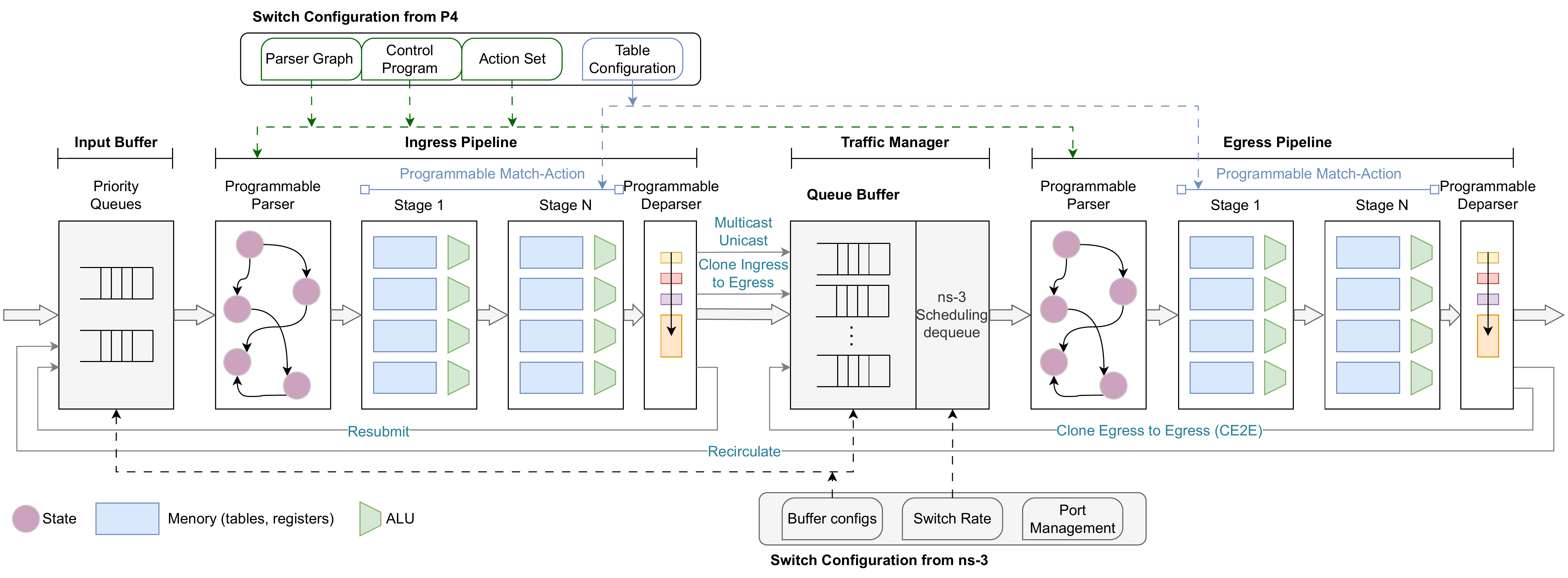}
  \caption{Internal structure of the proposed P4 switch model implemented in ns-3, following the PSA architecture. The figure illustrates two key aspects: the packet processing pipeline configured via the P4 program, and the buffer management and scheduling mechanisms configured within ns-3.}
  \label{fig:ns3_queue_design}
  \Description{Internal structure of the proposed P4 switch model implemented in ns-3, following the PSA architecture. The figure illustrates two key aspects: the packet processing pipeline configured via the P4 program, and the buffer management and scheduling mechanisms configured within ns-3.}
\end{figure*}

After identifying these key aspects, we introduce our implementation for P4 packets processing approach. In Section \ref{subsec:p4andbmv2}, we discussed the three bmv2 architectures: V1model, PSA, and PNA. Given that the V1model and PNA share structural similarities with PSA, we use the PSA as the basis for our analysis in this section. The packet processing path follows the PSA structure, including the ingress and egress pipelines, and each pipeline includes the programmable parser, match-action, and deparser. This programmable processing pipeline could be configured with P4 and flow table configuration.

Based on the P4 processing pipeline, we introduced input and queue buffers into ns-3, while omitting the transmit buffer. This buffer design has been adapted and modified from the bmv2, as illustrated by the gray boxes in Figure \ref{fig:ns3_queue_design}.
The input buffer is a priority queue with two subqueues, external packets are assigned lower priority as default, while resubmitted or recirculated internal packets are prioritized, ensuring that internal packets are processed first. 
The queue buffer contains up to eight virtual queues per port, which are dequeued in strict priority order. 
The transmit buffer is omitted in Figure \ref{fig:ns3_queue_design}. In the switch model, the transmit buffer would serve only as a temporary staging area for sending packets to the network, without adding any additional functionality.
The output port of the switch could utilize queue discipline (\textit{QueueDisc} from the ns-3 traffic-control module), which already manages packet queuing and scheduling for sending out. Following Occam's razor, we avoid unnecessary duplication and do not implement a separate transmit buffer within the switch.
In addition to buffering the packet, the buffers in Figure \ref{fig:ns3_queue_design} also support the execution of primitive actions. Represented by the gray arrows in Figure \ref{fig:ns3_queue_design}, such as recirculate, resubmit, and clone operations. The packets will be added to the queue buffer for the primitive action processing.


\subsection{Scheduling Model with Time }

In ns-3, virtual time does not advance automatically unless explicitly controlled by scheduled events or time-based mechanisms. This means that if no events are scheduled to trigger time advancement, the simulation virtual time effectively remains paused at the last processed event. Section \ref{sec:related_work} discusses the P4 switch time representation missing issues in NS4.
To address this limitation, we introduce event-based periodic scheduling, which generates events based on the switch rate to enable virtual state and time changes within the switch. This scheduling mechanism is indicated in Figure \ref{fig:ns3_queue_design}. The switch creates events at a frequency equal to its maximum packet forwarding rate. Consequently, the dequeue attempt advances the simulation time by 1 / rate seconds. 
In normal cases, each buffer should have its scheduling method and dequeue rate. However, in a switch with two queues, a single packet would be scheduled two times in one switch, significantly reducing simulation efficiency. To mitigate this, only at the queue buffer level apply the scheduling. This approach achieves one event per packet per switch, enhancing efficiency by reducing redundant scheduling overhead and minimizing processing delays.

In queue buffer, each virtual queue operates using a token-bucket-like mechanism, where external packets are assigned a label indicating their expected dequeue time. Packets can only be dequeued once this time is reached. 
This expected dequeue time is based on the packet processing rate of this queue. Packets are dequeued based on this time information, ensuring that each queue maintains its actual dequeue rate, which can be dynamically adjusted without modifying the switch's scheduling rate. The switch can allow users to modify queue rates and other configurations at runtime. Instead of using real machine time, we utilize ns-3's simulation virtual time to assign timestamps to packets and determine their dequeue time. 
With this design, the packet queuing latency is influenced by several factors, including the overall traffic arrival rate, the dequeue rates of individual virtual queues and the total system, the current queue lengths, and the priority distribution of packets.

By scheduling method, we introduce the time representation, but this time representation differs from conventional time models, which typically evaluate simulation time delays based on the system’s current state and parameters, and then execute functions after a fixed delay. In contrast, this model attempts function execution through polling, using the success of such attempts to determine the simulation time progression. The advantage of this approach is that it avoids the need for precise mathematical modeling and analytical solutions of the system. However, it introduces idle polling overhead.

Once scheduling is enabled in the switch model, it may need to retrieve the virtual time information from the simulator of ns-3. The V1Model introduces more timing-related requirements compared to PSA and PNA, we base our discussion on the V1Model.
The V1model requires \textit{ingress\allowbreak\_global\allowbreak\_timestamp}, \textit{egress\allowbreak\_global\allowbreak\_timestamp} for intrinsic metadata and \textit{enq\allowbreak\_timestamp}, \textit{deq\_timed\allowbreak elta} for queueing metadata.
Intrinsic metadata captures different phases of packet processing while queueing metadata reflects queue-related information. 
The event-based scheduling is implemented in a queue buffer between ingress and egress. As a result, each packet's ingress and egress timestamps include an increment equal to the queuing time. The intrinsic metadata \textit{egress\allowbreak\_global\allowbreak\_timestamp}, the queueing metadata \textit{enq\allowbreak\_timestamp}, and their difference \textit{deq\allowbreak\_timedelta} are correctly assigned with different value, ensuring precise time tracking throughout the switch operation.
The time interval between \textit{ingress\allowbreak\_global\allowbreak\_timestamp} and \textit{enq\allowbreak\_timestamp} represents the ingress processing time, including match-action execution. In the absence of hardware testing data, we currently assume these two timestamps are identical. Future refinements can incorporate hardware based processing delay models to improve accuracy.
In the PSA architecture, \textit{ingress\allowbreak\_timestamp} and \textit{egress\allowbreak\_timestamp} are required, while PNA only requires a timestamp for packets entering processing. The internal switch time representation used in the V1model timestamp implementation can effectively meet the timing requirements of both PSA and PNA.

\subsection{P4 Custom Header}

As mentioned in Section \ref{sec:related_work}, supporting the addition of custom headers in user-defined simulation scripts is essential for the rapid verification of P4 protocols and the implementation of specialized functions. However, ns-3 currently lacks a dedicated mechanism to define and simulate such custom headers, making it challenging to extend protocol functionality dynamically.
The packet's header processing in ns-3 follows a layered approach. The transport layer adds the UDP/TCP header before the payload, including port numbers and sequence details. The network layer incorporates an IPv4/IPv6 header with source and destination IP addresses, while the data link layer adds an Ethernet header containing MAC addresses and EtherType. Finally, the fully encapsulated packet is forwarded by the net devices over the network.
To streamline and standardize custom header definitions, we introduce an abstraction that categorizes headers based on their corresponding network layer. A custom header can be applied at the data link layer (\textit{LAYER\_2}), network layer (\textit{LAYER\_3}), or transport layer (\textit{LAYER\_4}). Users can control header placement by specifying one of three modes: \textit{ADD\_BEFORE} (insert before an existing header), \textit{ADD\_AFTER} (insert after an existing header), or \textit{REPLACE} (substitute an existing header).
This approach enhances flexibility in modifying packet structures, facilitating protocol experimentation, and enabling the rapid prototyping of novel networking concepts in ns-3 simulations.

\begin{figure}[ht]
    \centering
    \includegraphics[width=\linewidth]{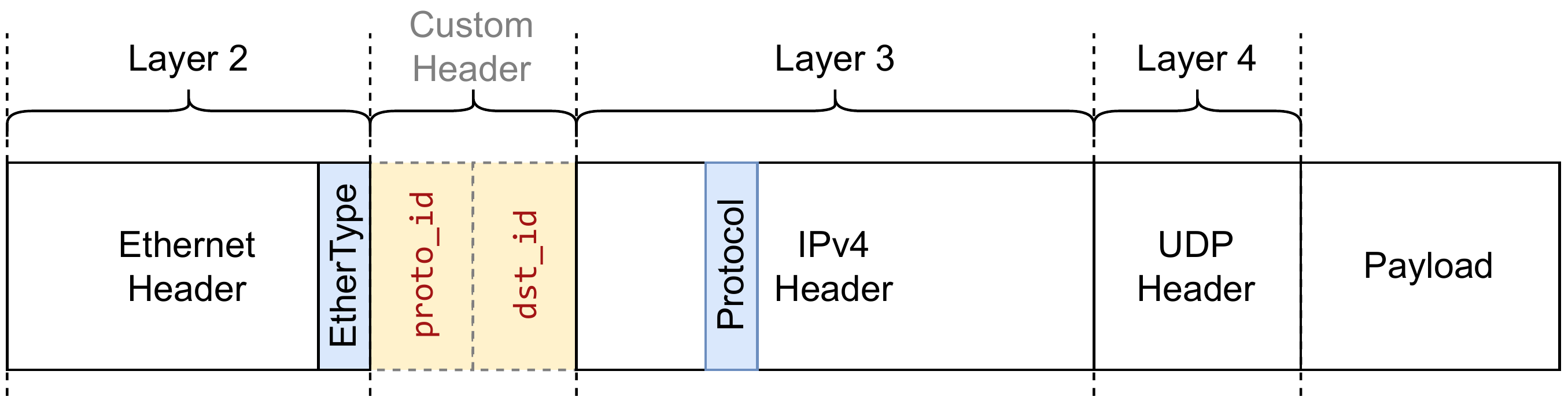}
    \caption{Example of the packet with tunnel custom header.}
    \label{fig:header_udp}
    \Description{A packet structure diagram illustrating multiple layers. It consists of an Ethernet header (Layer 2), a custom header with fields proto\_id and dst\_id, an IPv4 header (Layer 3), a UDP header (Layer 4), and a payload section. The custom header is inserted between the Ethernet and IPv4 headers.}
\end{figure}

To illustrate this approach, Figure \ref{fig:header_udp} shows how a custom tunnel header is inserted between the Ethernet and IPv4 headers. This placement is defined using \textit{LAYER\_3} and \textit{ADD\_BEFORE}. Listing \ref{lst:tunnel-header} provides a code implementation of this approach, where the tunnel header includes fields for the protocol ID (\textit{proto\_id}) and destination ID (\textit{dst\_id}), each 16 bits in length.
To ensure correct parsing and debugging, the \textit{proto\_id} field must be included in the custom header to indicate the type of the subsequent header. If omitted, the packet parser may fail to correctly interpret the following protocol headers, leading to parsing errors in both software and hardware implementations.


After the custom header is defined, its proper processing function must be implemented on the net device.
To enable customization of Layer 2-4 headers, we introduce an approach for dynamically modifying headers at the lowest level of the net device. This allows researchers to insert, modify, or remove headers while ensuring that upper-layer protocols such as ARP, IPv4, and ICMP remain unaffected. 
On the sender-side net device, headers only need to be assembled in order. However, on the receiver-side net device, since no sequence information is available, it must iterate through each header's protocol field and use the protocol number to determine the next parsing step or reception.  
This design requires all participating network hosts to install these net devices to ensure proper recognition and processing of packet headers.

\begin{lstlisting}[style=customcpp, label={lst:tunnel-header}, caption={Example of tunnel custom header definition.}]
  CustomHeader myTunnelHeader;
  myTunnelHeader.SetLayer (HeaderLayer::LAYER_3);
  myTunnelHeader.SetOperator (ADD_BEFORE);
  myTunnelHeader.AddField ("proto_id", 16); // 16 bits
  myTunnelHeader.AddField ("dst_id", 16); // 16 bits
  myTunnelHeader.SetField ("proto_id", 0x0800); // IPv4
  myTunnelHeader.SetField ("dst_id", 0x22);
\end{lstlisting}

\section{P4sim Implementation}
\label{sec:moduleimplementation}

\begin{figure*}[htbp]
    \centering
    \includegraphics[width=\textwidth]{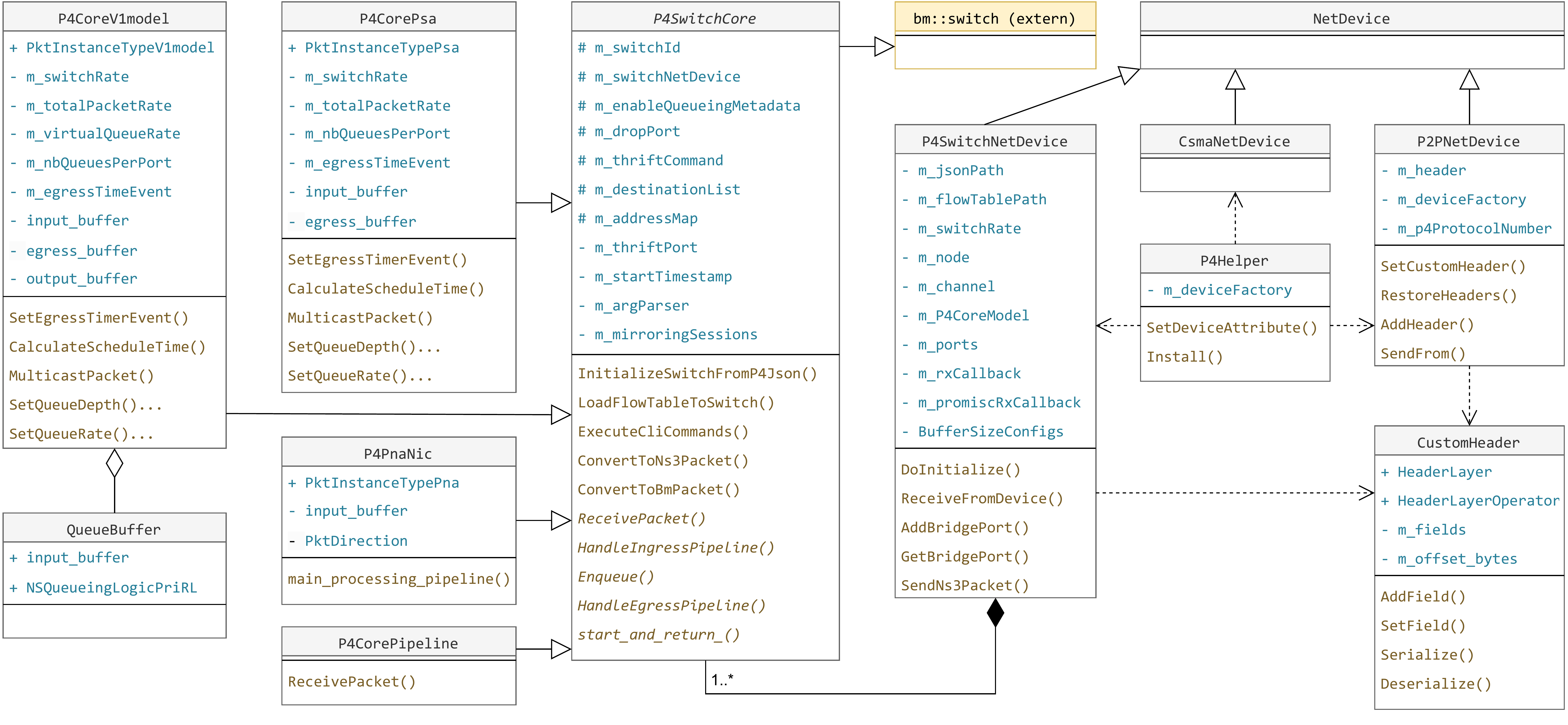}
    \caption{UML diagram of the P4 switch implementation, showing only the important newly added classes, functions, and variables.}
    \label{fig:uml}
    \Description{A UML diagram representing the P4 switch implementation. It highlights the newly added classes, functions, and variables, illustrating their relationships and interactions within the system.}
\end{figure*}

In this section, based on the approach as mentioned in section \ref{sec:ns3p4simulator}, we examine the fundamental mechanisms underlying the implementation of P4 switch, with a particular emphasis on key classes involved in packet processing. This analysis includes an overview of the existing classes in the ns-3 library and the introduction of new classes aligned with the Unified Modeling Language (UML) diagram presented in Figure \ref{fig:uml}. It concisely illustrates the essential relationships, dependencies, and attributes of the core classes that define our approach. The key classes involved are outlined below, highlighting their roles, relationships, and dependencies in our implementation.

\subsection{P4SwitchNetDevice}


The \textit{P4SwitchNetDevice} class serves as the underlying switch component in ns-3, coordinating ns-3-specific elements such as ports and hosts. 
It extends the \textit{NetDevice} class and is designed to manage multiple network ports. These ports can be instances of either \textit{PointToPointNetDevice} or \textit{CsmaNetDevice}, providing flexibility in network topology configuration. Additionally, it supports dynamic port management, allowing developers to add or remove ports as needed.
The primary functions of \textit{P4SwitchNetDevice} include initializing \textit{P4SwitchCore} for packet processing, packet reception and transmission. The P4-based packet processing is executed within \textit{P4SwitchCore}. In the case of a V1model architecture, each \textit{P4SwitchNetDevice} is equipped with a \textit{P4CoreV1model} processing module. 
To facilitate deployment, the \textit{P4Helper} utility simplifies the installation and configuration of \textit{P4SwitchNetDevice}, enabling efficient integration within network simulation environments. 
This switch device can be configured with \textit{CsmaNetDevice} as a port, the channel should be set to a CSMA channel, with proper handling of ARP packets. Additionally, it can also incorporate \textit{CustomP2PNetDevice} in Figure \ref{fig:uml} as a port, where the corresponding channel should be the \textit{P4P2PChannel} channel. 


\subsection{P4SwitchCore}


The \textit{P4SwitchCore} class serves as the core of the switch, managing P4-specific implementations such as packet processing and forwarding logic, it handles key functionalities including queue management, scheduling, and runtime reconfigurations.
\textit{P4SwitchCore} module extends the external \textit{bm::switch} and implements its functionality within the ns-3 network simulation framework. 
The \textit{P4SwitchCore} handles external packets received from the \textit{P4SwitchNetDevice}, processes them, and forwards them through the selected output port.
It has several subclass implementations, each corresponding to a specific architecture, such as V1model from \textit{P4CoreV1model}, PSA from \textit{P4CorePsa}, and PNA from \textit{P4PnaNic}. 
It is initialized using a compiled P4JSON file, which contains configuration and mapping information for various components defined in the P4 language. This includes the parser, tables, and actions for ingress and egress, deparser, as well as configurations for registers and other resources. After initialization, the system enables the Thrift port, allowing flow table entries to be configured dynamically through the Thrift interface.


\subsection{CustomHeader and CustomP2PNetDevice}

The \textit{CustomHeader} class implements user-defined headers, allowing users to define a header and bind it to \textit{CustomP2PNetDevice} (\textit{CPDevice}), which then provides support for the custom header. 
Inheriting from the \textit{Header} class, \textit{CustomHeader} supports arbitrary custom fields and can be placed at the L2, L3, or L4 layers. Additionally, it allows setting specific values for corresponding fields within the data flow header. 
To support this header on both the sender and receiver sides, we modified the \textit{PointToPointNetDevice} to \textit{CPDevice} and introduced functions for custom header insertion and extraction. The original Point-to-Point Protocol (PPP) in the default net device has been replaced with the Ethernet II protocol. Notably, the Frame Check Sequence (FCS) is omitted from the transmitted packets. This implies that the \textit{CPDevice} functions as an Ethernet port. Additionally, we introduce a simple \textit{P4P2PChannel} as its full-duplex high-speed connection, which is not configured with any error model.

Our implementation leverages the \textit{CPDevice} on the sender side to apply custom header modifications before packet transmission. On the receiver side, the \textit{CPDevice} interprets and processes these headers, ensuring that the changes remain transparent to higher-layer protocols. This enables seamless integration with existing network stacks while allowing network switches to process, forward, or modify packets based on researcher-defined P4 configurations. Researchers only need to define the custom header format and configure the \textit{CPDevice} accordingly, without requiring modifications to ns-3’s core transport and network-layer protocols.

To enable \textit{CPDevice} to add or retain the original application-generated traffic automatically, we design a mechanism that dynamically determines header insertion based on destination port ranges. 
Traffic with UDP or TCP destination port numbers between $10,000$ and $12,000$ 
is augmented with a custom header, while traffic outside this range is transmitted without modification. The port numbers of TCP or UDP traffic can be easily configured in ns-3 using the application helper. This approach ensures seamless integration of custom headers while preserving standard packet processing for regular traffic. This automation reduces the need for manual configuration, making it easier for researchers to validate P4-based header modifications in ns-3 simulations.


\section{Evaluation}
\label{sec:evaluation}

This section presents a series of experimental evaluations to validate the implementation of P4sim and to address the requirements R1–R4 outlined in Section \ref{sec:introduction}. R1 is validated by evaluating the behavior and performance of the P4 switch, along with demonstrating its functionality in a load balancing scenario. R3 is verified through a basic tunneling experiment using a custom tunnel header. Additionally, for R2 we include IPv4 forwarding examples for the V1Model, PSA, and PNA architectures in open-source repository. The implementation supports both the waf build system (for ns-3.35 and earlier) and the CMake build system (for ns-3.39 and earlier), compiled with C++17. The test environment is deployed using a virtual machine provisioned through a Vagrant script. All experiments and tests are conducted within this virtual environment.

\subsection{Performance}

To assess the correctness and performance of the module, we conduct tests using the basic \textit{ipv4\_forward} communication scenario based on the V1Model architecture. Since the P4sim module is implemented in ns-3 by integrating the bmv2, we choose bmv2 with Mininet, and ns-3 simulations without P4 as the benchmark software platform. The performance and simulation efficiency are evaluated by throughput and simulation wall-clock time. Each experiment is conducted 10 times.
The network topology consists of a single switch connected to two hosts. The hosts exchange data packets, and the switch is responsible for forwarding them based on their IPv4 addresses. Both CSMA and P2P channels are available for P4sim. However, since the ns-3 bridge does not support P2P channels, this evaluation uses a CSMA channel. The P2P channel evaluation will be shown in Section \ref{subsec:loadbalancing}. 

\begin{figure}[ht]
    \centering
    \includegraphics[width=\linewidth]{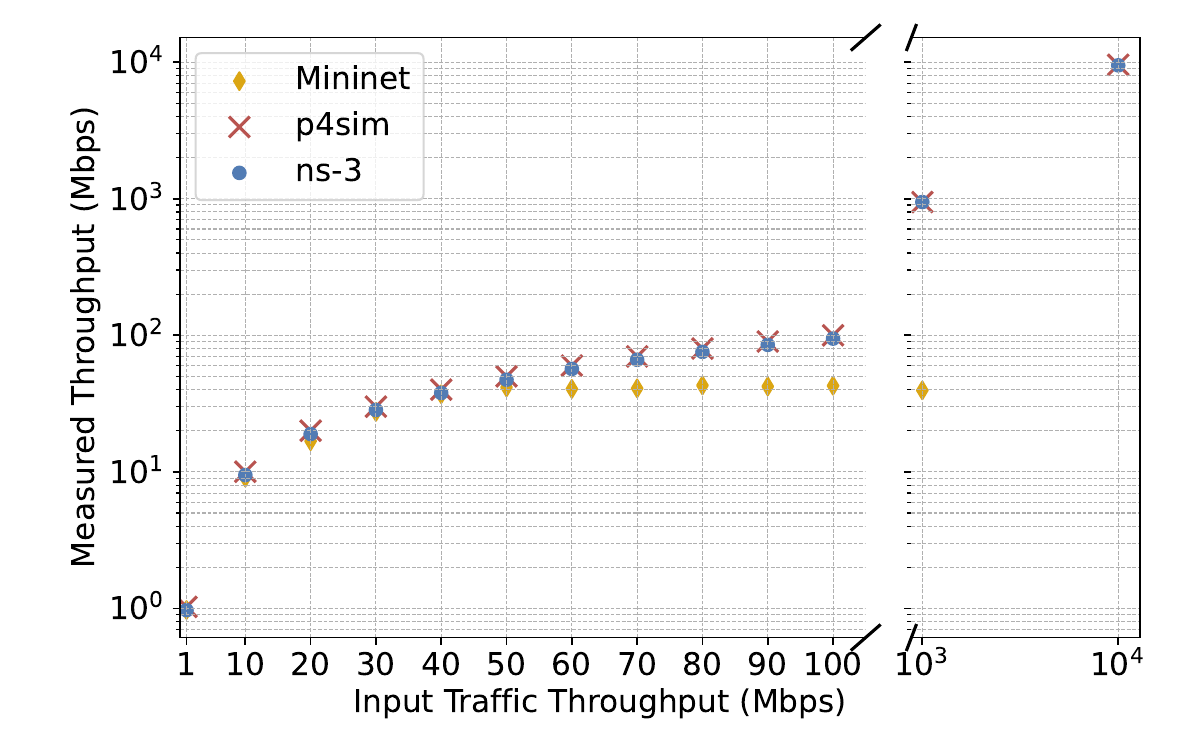}
    \caption{Evaluation of network performance through packet forwarding.}
    \label{fig:network_throughput_comparison}
    \Description{The network throughput.}
\end{figure}

Figure \ref{fig:network_throughput_comparison} presents the averaged measured throughput under different configured bandwidths. The x-axis represents the expected transmission rate, which corresponds to the configured sending rate rather than the actual link capacity. The evaluation covers a range from $1 Mbps$ to $100 Mbps$ in $10 Mbps$ intervals, as well as $1000 Mbps$ and $10000 Mbps$. The y-axis shows the actual measured throughput at the receiver host, plotted on a logarithmic scale to better visualize variations across different bandwidths. In Mininet, the throughput begins to saturate at around $43 Mbps$ as the input traffic rate, indicating a performance bottleneck, beyond which increasing the input traffic rate does not significantly improve the actual throughput. In contrast, P4sim and the ns-3 bridge exhibit similar throughput performance with the input traffic rate. 
This highlights the simulator’s advantage in measuring high-speed networks and confirms that P4sim P4 switch and ns-3 bridge-based simulations offer reliable performance in different network scenarios.

\begin{figure}[ht]
    \centering
    \includegraphics[width=\linewidth]{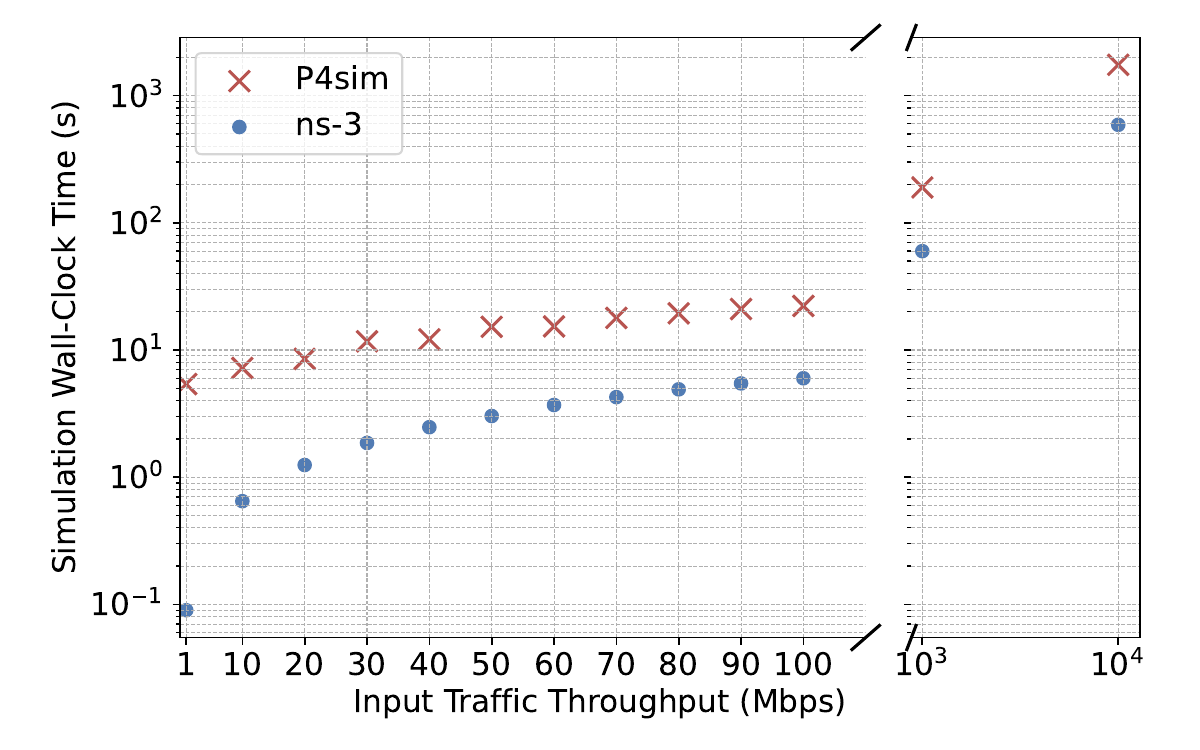}
    \caption{Evaluation of simulation efficiency based on wall-clock execution time.}
    \label{fig:network_simulation_time}
    \Description{The time usage for simulation.}
\end{figure}

We further compare the simulation wall-clock time between ns-3 and P4sim for a simulated duration of 1 second, as shown in Figure \ref{fig:network_simulation_time}. The y-axis represents the simulation time in seconds on a logarithmic scale. 
The figure shows that P4sim consistently requires more simulation time than ns-3 under the same network simulation conditions. Notably, despite the absolute difference in execution time, the two curves exhibit a nearly parallel trend across the throughput range. In a log-scale plot, such parallelism indicates that both P4sim and ns-3 share a similar growth rate in simulation wall-clock time as traffic increases. The vertical offset between the two curves remains nearly constant at approximately $5 dB$, indicating that P4sim's execution time is consistently about $3.16$ times that of ns-3 across different input traffic levels.
This stable ratio suggests that although both simulators exhibit similar growth trends in simulation time, P4sim consistently incurs a significantly higher wall-clock execution time. This discrepancy is primarily due to the initialization overhead in P4sim, which includes the startup procedures, flow table deployment, and, most notably the increased packet processing time introduced by the P4 programmable pipeline.

\subsection{Use Cases}


\subsubsection{Basic Tunneling}

In this experiment, we extend the functionality of a basic IP router by incorporating a custom tunneling mechanism designed to support encapsulation-based forwarding. Tunneling is a common technique used in network virtualization, overlay networks, or service chaining, where original packets are encapsulated within an additional header that guides intermediate forwarding decisions. 
The implementation introduces a new encapsulation header, \textit{myTunnel\_t}, which carries two key fields: a protocol ID (\textit{proto\_id}) to indicate the type of payload encapsulated, and a destination ID (\textit{dst\_id}) used to determine the forwarding path. The structure of this header is detailed in Listing \ref{lst:tunnel-header}.
The P4 switch is modified to parse this new header type, extract its fields based on the Ethernet protocol type (we assume $0x1212$), and apply different forwarding behaviors accordingly. A new \textit{myTunnel\allowbreak\_exact} table is implemented to perform exact matching on the \textit{dst\_id} field, triggering a forwarding action based on the matched value.
Furthermore, the ingress pipeline is updated to prioritize tunnel-based forwarding for packets containing the myTunnel\_t header, while still preserving standard IPv4 forwarding behavior for normal traffic. This hybrid approach allows the switch to seamlessly support both encapsulated and non-encapsulated flows.The deparser ensures the correct reconstruction of the packet by emitting headers in the expected order—Ethernet, followed by myTunnel\_t, and then the inner IP header—ensuring end-to-end compatibility across network devices.

\begin{figure}[ht]
    \centering
    \includegraphics[width=\linewidth]{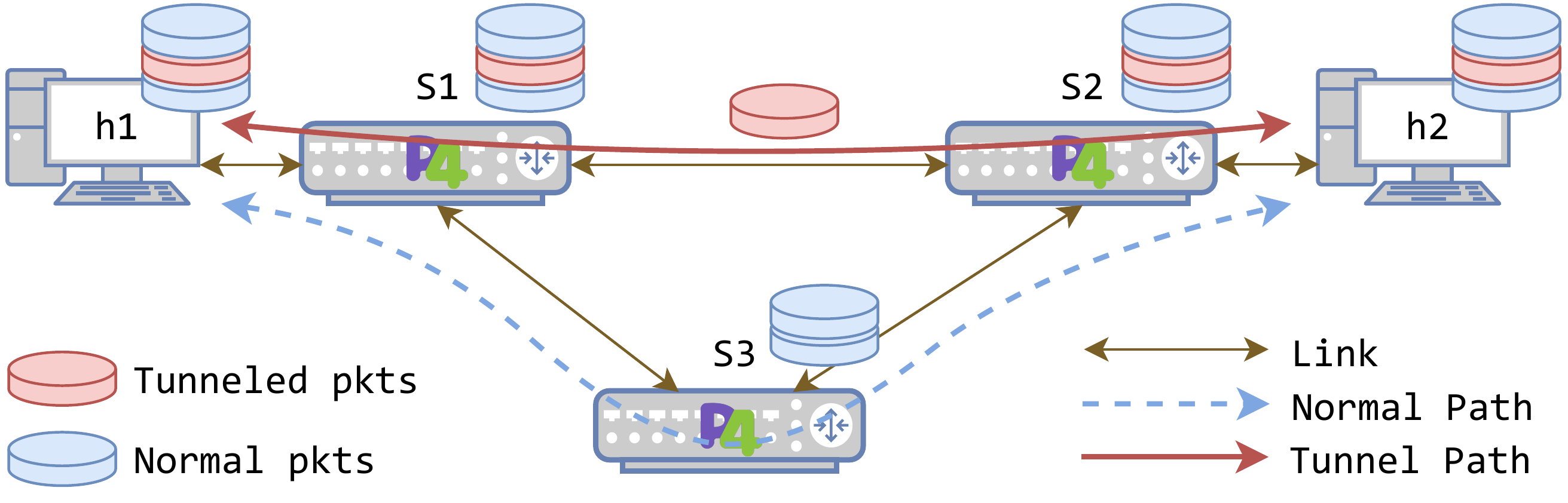}
    \caption{Data transmission from h1 to h2 in a P4 network: normal and tunnel traffic \cite{p4lang/tutorials/exercises/basic_tunnel}.}
    \label{fig:tunneling}
    \Description{A network diagram illustrating data transmission from host h1 to host h2 in a P4 network. The network consists of three P4 switches (S1, S2, and S3) interconnected by links. Normal packets are shown following a blue dashed normal path, while tunneled packets follow a red solid tunnel path. Each host and switch is depicted with corresponding buffer representations for normal and tunneled packets.}
\end{figure}

In this network scenario, h1 transmits tunneled traffic at $10 Mbps$ with $1,250 pps$ and normal traffic at $40 Mbps$ with $5,000 pps$ to h2.
The normal traffic follows the path S1, S3, S2, h2, while the tunneled traffic takes a direct route through S1, S2, h2. 
The characteristics of the packet can be captured as \textit{pcap} files and analyzed in Wireshark. However, custom headers may not be automatically parsed correctly. If Wireshark fails to recognize the header, throughput calculations in $bps$ will be inaccurate since the header is included in the data. Therefore, we use packet count ($pps$) as the evaluation metric.
For the test results, S1 forwards approximately $5,000 pps$ to S3 and $1,250 pps$ to S2, while S2 receives a total average of $6,250 pps$. 
This result indicates that the tunnel header functions as intended, enabling the configured traffic to be transmitted and forwarded through the tunnel at the expected rate.

\subsubsection{Load Balancing}
\label{subsec:loadbalancing}

\begin{figure}[ht]
    \centering
    \includegraphics[width=\linewidth]{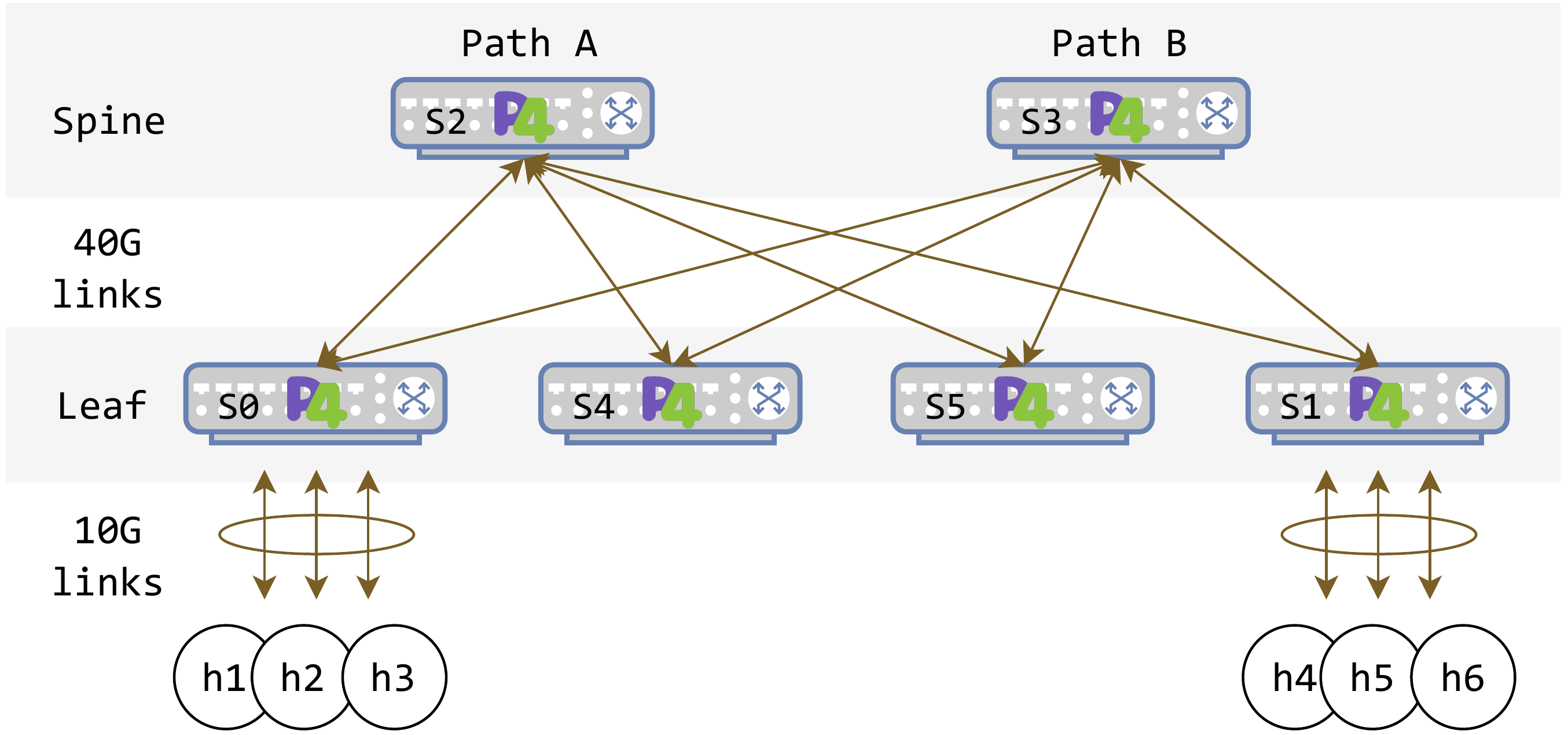}
    \caption{The spine-leaf network topology for simulation ECMP load balancing \cite{p4lang/tutorials/exercises/loadbalancing}.}
    \label{fig:loadbalancing}
    \Description{A network topology diagram illustrating an ECMP load balancing setup using a spine-leaf architecture. The topology consists of two spine P4 switches (S2, S3) connected to four leaf P4 switches (S0, S4, S5, S1) via 40G links. Hosts h1, h2, and h3 are connected to leaf switch S0 via 10G links, while hosts h4, h5, and h6 are connected to leaf switch S1 via 10G links. Arrows indicate bidirectional connections between switches and hosts.}
\end{figure}

In this experiment, we demonstrate a load balancing use case, as shown in Figure \ref{fig:loadbalancing}. The network adopts a spine-leaf architecture, which is widely used in modern data center designs due to its scalability, high availability, and its ability to efficiently handle large volumes of East-West traffic.
Load balancing is essential to efficiently distribute traffic across multiple paths, alleviate congestion, and improve overall network performance. In this architecture, leaf switches connect directly to servers, providing access to network hosts, while spine switches interconnect all leaf switches, ensuring multiple redundant paths for data transmission. 
To achieve load balancing, all switches are configured with static hash-based Equal-Cost Multi-Path (ECMP) routing. Each switch determines the forwarding path by applying a 5-tuple hash function to incoming packets, ensuring traffic is distributed across available links in a balanced way.

In the setup shown in Figure \ref{fig:loadbalancing}, host h1 generates $1,000$ UDP flows, each assigned to a different destination port. Each flow operates at $10 \space Mbps$, resulting in a total maximum transmission rate of $10 \space Gbps$, which matches the single-link capacity. 
Traffic generation follows an On-Off pattern using the \textit{OnOffApplication}. The on-and-off periods are exponentially distributed with mean values of 2 seconds and 1 second. 
The generated traffic is transmitted over the modified P2P channel and ultimately reaches the destination, host h4. We have evaluated the traffic throughput at key nodes. The input traffic and received traffic are measured at the leaf nodes S0 and S1, representing the total traffic entering and exiting the system. Path A and Path B correspond to the two forwarding paths through the spine nodes S2 and S3, where traffic is measured.

\begin{figure}[ht]
    \centering
    \includegraphics[width=\linewidth]{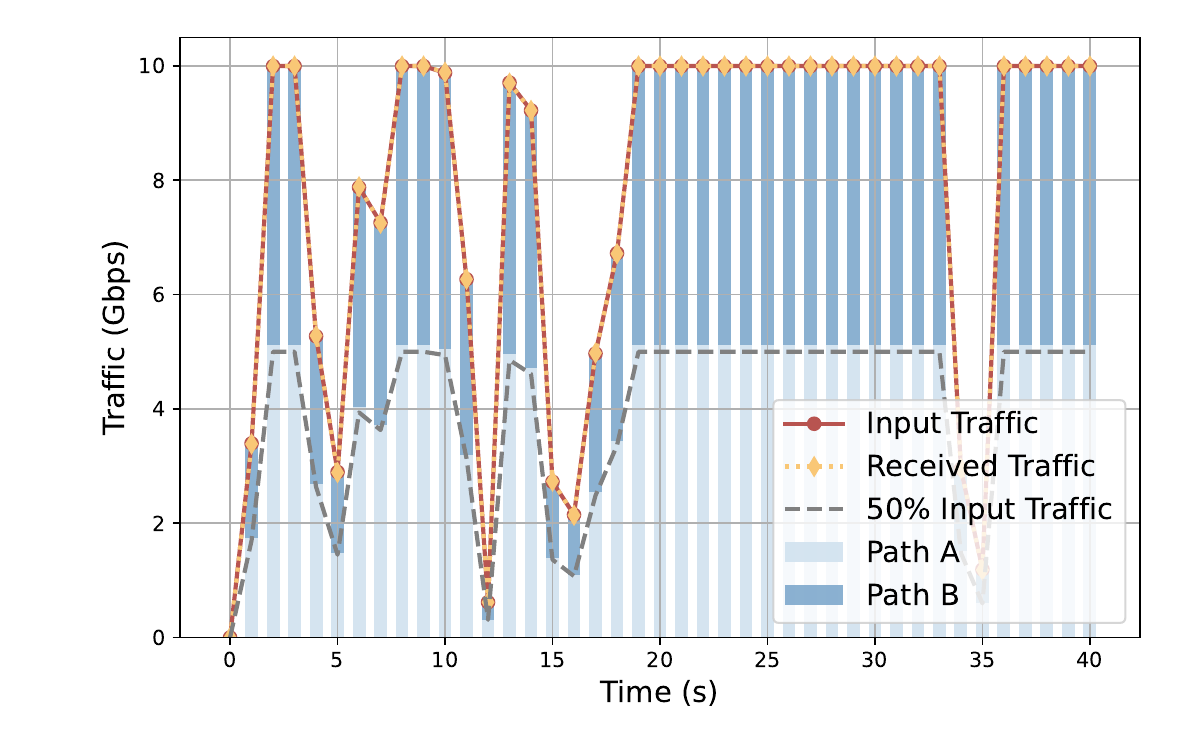}
    \caption{System throughput over time.}
    \label{fig:loadbalancingthroughput}
    \Description{This figure illustrates the system throughput over time, showing how the input traffic is distributed between two paths (A and B) and received at the destination. The stacked bars represent paths A and B, while the solid and dashed lines indicate input and received traffic. A gray dashed line represents 50 percent of the input traffic for load balancing comparison.}
\end{figure}

The simulation result of load balancing is shown in Figure \ref{fig:loadbalancingthroughput}. 
The red line represents the input traffic from host h1, while the yellow dashed line indicates the received throughput at host h4. The light blue bar (Path A) and dark blue bar (Path B) show the throughput measured at switches S2 and S3 in the specific time.  
To better illustrate the load balancing performance, a gray dashed line is included, representing 50\% of the input traffic, serving as a reference to indicate whether the traffic is evenly distributed between the two paths. 
During the simulation period from 0 to 40 seconds, the input traffic matches the received traffic and fluctuates within the range of $0$ to $10 Gbps$. The combined throughput of Path A and Path B equals the input (or received) traffic. Although the traffic is nearly balanced between the two paths, Path B consistently carries slightly more traffic than Path A, rather than an exact $50/50$ split. On average, over the 40-second simulation, the throughput ratio of Path A to Path B is approximately $1.0458$.
These results demonstrate that the P4-based load balancing implementation in P4sim functions correctly and effectively. The total received throughput matches the input traffic. The near-even distribution of traffic across both paths confirms the effectiveness of the hash-based ECMP strategy. The slight imbalance, reflected by the ratio of 1.0458, is likely due to the limited number of flows (only 1,000), which is insufficient to achieve perfectly uniform distribution using a 5-tuple hash function.
Overall, the experiment verifies that P4sim supports programmable data plane behavior with reliable performance in a realistic load balancing scenario.

\section{Conclusion and Future Work}
\label{sec:conclusion}

This paper introduced P4sim, an enhanced P4-driven simulation framework addressing the limitations of existing platforms in high-speed, large-scale network simulations. Built on NS4, it aligns with ns-3’s modular design while improving queue modeling, time scheduling, and expanding P4 architecture support.
P4sim integrates V1model, PSA, and PNA architectures, enabling more flexible and diverse pipeline behaviors. Additionally, its protocol-independent custom header support ensures seamless interaction between P4-enabled hosts and switches, enhancing flexibility and ease of integration. The framework also maintains tight integration with ns-3, supporting modular builds with CMake and Waf.
Through performance and accuracy evaluations, we validated P4sim’s efficiency and correctness, demonstrating its practicality with Basic Tunneling and Load Balancing use cases. 

\balance

In the current switch model, NetDevice at each port does not provide feedback to the switch regarding packet transmission status. Once the switch processes a packet, it hands it over to the port NetDevice, which then transmits it to the Channel. However, if the Channel is congested or unable to immediately accommodate the packet, it is buffered in the port’s queue. Since the switch lacks visibility into this queue, P4 scripts are unable to manage congestion or perform scheduling based on queue occupancy. 
Future research could explore mechanisms for integrating queue state awareness into the switch model, enabling P4-based congestion control and scheduling strategies. This enhancement would allow for more accurate simulation of network congestion and improve traffic management capabilities within the simulator.

\begin{acks}
\small
Funded by the German Research Foundation (DFG, Deutsche Forschungsgemeinschaft) as part of Germany’s Excellence Strategy – EXC 2050/1 – Project ID 390696704 – Cluster of Excellence “Centre for Tactile Internet with Human-in-the-Loop” (CeTI) of Technische Universität Dresden.
\end{acks}

\bibliographystyle{ACM-Reference-Format}




\end{document}